\documentstyle[prl,aps,floats]{revtex}

\begin{document}
\draft
\def\overlay#1#2{\setbox0=\hbox{#1}\setbox1=\hbox to \wd0{\hss #2\hss}#1
-2\wd0\copy1}
\twocolumn[\hsize\textwidth\columnwidth\hsize\csname@twocolumnfalse\endcsname

\title{Quantum-Non-Demolition Endoscopic Tomography}

\author{Mauro Fortunato\cite{mau} and Paolo Tombesi}
\address{Dipartimento di Matematica e Fisica, Universit\`a di
        Camerino, I-62032 Camerino, Italy \\ and Istituto Nazionale per la
        Fisica della Materia, Unit\`a di Camerino}
\author{Wolfgang~P.~Schleich}
\address{Abteilung f\"ur Quantenphysik, Universit\"at Ulm, D-89069 Ulm,
         Germany}
\date{Received \today}
\maketitle
\begin{abstract}
We present a new indirect method to measure the quantum state of a
single mode of the electromagnetic field in a cavity. Our proposal
combines the idea of (endoscopic) probing and that of tomography in
the sense that the signal field is coupled  via a quantum-non-demolition
Hamiltonian to a meter field on which then quantum state tomography
is performed using balanced homodyne detection. This technique provides
full information about the signal state. We also discuss the influence
of the measurement of the meter on the signal field. 
\end{abstract}
\vskip2pc]
\narrowtext

\section{Introduction: State Measurement}
\label{intro}

The question of how to measure the quantum state of a single mode of the
electromagnetic field in a cavity has recently attracted a great deal 
of attention~\cite{kn:spec}, as it determines the feasibility of the
{\em measurement} of the state of a genuine quantum system~\cite{kn:disc}.
Several proposals have been made in the last few years to answer this
question. Among them, quantum-state endoscopy~\cite{kn:bard} and
quantum-optical homodyne state-tomography~\cite{kn:tomo} are two
notable examples.
The former proposal makes use of a beam of two-level
atoms---sent with controlled speed through the cavity---to infer 
the properties of the field state inside the cavity.
On the other hand, optical homodyne tomography is a method for obtaining
the Wigner function (or, more generally, the matrix
elements of the density operator in some representation) of the electromagnetic
field. It therefore consists of an ensemble of repeated measurements of one
quadrature operator for different phases relative to the local oscillator of
the homodyne detector. The major drawback of the first method is the 
low detection efficiency for atoms, whereas in the second one one has
to couple the field out of the resonator.

In the present paper we propose to couple the field via a
quantum-non-demolition (QND) interaction~\cite{kn:sum} to a meter field on
which we then perform tomography using a balanced homodyne detector.
In this way we combine the idea of probing, that is doing endoscopy on the 
field without taking it out of the cavity, and the tool of tomography and
arrive at the method of endoscopic quantum state tomography.
In contrast to the method of quantum state tomography~\cite{kn:tomo}
based on homodyne detection, the present technique does not couple
the signal field out of the resonator.

\section{Our Model}
\label{hami}

Let us assume that the electromagnetic field we want to probe (the 
signal mode) is in a pure quantum state and neglect dissipation.
We note, however, that our method applies as well 
to a field described by a density operator.
In order to measure the signal field
we couple it to a meter field. Both fields are then coupled to 
a pump field. This coupling leads us to a quantum-non-demolition Hamiltonian
describing the interaction between the signal and the meter mode.

Our model starts from the Hamiltonian 
\begin{equation}
\hat {H} \equiv i \hbar \chi [\hat {a}_s^\dagger \hat {a}_m^\dagger \hat 
{a}_p - \hat {a}_s \hat {a}_m \hat {a}_p^\dagger] + i\hbar \sigma [\hat {a}_s
^\dagger \hat {a}_m - \hat {a}_s \hat {a}_m^\dagger]\;,
\label{eq:hami}
\end{equation}
where $\hat{a}_s (\hat{a}_s^\dagger)$, $\hat {a}_m (\hat{a}_m^\dagger)$, and
$\hat{a}_p (\hat{a}_p^\dagger)$ represent the annihilation (creation) operators 
of the signal, meter, and pump field, respectively. The parameters $\chi$
and $\sigma$ measure the coupling between the three fields, and the meter and 
signal field, respectively. When the pump field is highly
excited we can describe it by a coherent state of amplitude $\alpha$ and 
phase $2\phi$, that is 
\begin{equation}
\hat{a}_p \simeq \alpha e^{2i\phi}\;. 
\label{eq:coh}
\end{equation}
If we
substitute the coherent state approximation, Eq.~(\ref{eq:coh}), into
the Hamiltonian, Eq.~(\ref{eq:hami}), after some algebra we obtain
\begin{equation}
\hat{H} = 2\hbar \sigma \hat X_s (\phi + \pi/2) \cdot \hat{X}_m (\phi)\;,
\label{eq:cohami}
\end{equation}
where we have arranged the strength $\alpha$ of the pump field such that
$\chi\alpha = \sigma$, and the quadrature operators are given by
\begin{equation}
\hat{X}_j (\theta) \equiv \frac{1}{\sqrt{2}} \left(\hat{a}_j e^{-i\theta} +
\hat{a}_j^\dagger e ^{i\theta}\right)
\end{equation}
of the signal $(j=s)$ and the meter $(j=m)$ mode at the angle $\theta$.

It is particularly interesting to note that due to the special choice
$\chi\alpha = \sigma$ we have been able to obtain an interaction between
the signal and the meter which couples the quadrature operator $\hat{X}_m (\phi)$ of the meter at 
phase angle 
$\phi$ to the out-of-phase quadrature operator $\hat{X}_s(\phi + \pi/2)$
of the signal~\cite{kn:als,kn:schl}. The present Hamiltonian is a 
genuine QND Hamiltonian. In the next sections we shall see
how such a Hamiltonian can be used to measure the quantum state
of the signal field. We note that according to the QND Hamiltonian
Eq.~(\ref{eq:cohami}) a (homodyne) measurement of the meter at a fixed
phase $\phi$ of the pump field yields information about the signal in the
orthogonal quadrature. By varying the phase $\phi$ of the pump field we can
probe all quadratures of the signal.

\section{Entangled State}
\label{enta}

The aim of the present section is to calculate the combined state
$|\Psi\rangle$ of signal and meter achieved after some interaction 
time according to the QND interaction Hamiltonian, Eq.~(\ref{eq:cohami}).

When we couple for an interaction time $t$ the signal and meter mode prepared
initially in the states 
$|\psi_s\rangle$ and $|\psi_m\rangle$ we find the quantum state
\begin{eqnarray}
|\Psi(t) \rangle &=& \exp (-i\hat{H}t/\hbar)|\psi_m\rangle|\psi_s\rangle
\nonumber \\
 & = & \exp[-2i\sigma t\hat{X}_s(\phi + \pi/2)
\hat{X}_m(\phi)]|\psi_m\rangle|\psi_s\rangle
\label{eq:evol}
\end{eqnarray}  
for the combined system.

We expand the initial signal state in
quadrature states $|X_s(\phi + \pi/2)\rangle$,
\begin{equation}
|\psi_s\rangle = \int\limits_{-\infty}^\infty dX_s\, \psi_s(X_s; \phi + \pi/2)
|X_s (\phi + \pi/2)\rangle\;.
\label{eq:exp}
\end{equation}
We stress that this representation and, in particular, the wave function
$\psi_s(X_s; \phi + \pi/2) \equiv \langle X_s(\phi + \pi/2)|\psi_s\rangle$
depend crucially on the angle $\theta_s$.

Combining Eqs.~(\ref{eq:evol}) and (\ref{eq:exp}), we may rewrite the
combined state as
\begin{eqnarray}
|\Psi(t) \rangle = \int\limits_{-\infty}^\infty & dX_s & \psi_s (X_s; \phi +
\pi/2)|X_s(\phi + \pi/2)\rangle
\nonumber \\
 & \times & \exp[-2i\sigma t X_s \hat{X}_m (\phi)]|\psi_m\rangle\;.
\label{eq:comb}
\end{eqnarray}

Expanding $|\psi_m\rangle$ in quadrature states $|X_m(\theta)\rangle$ of the
meter at the angle $\theta$, that is
\begin{equation}
|\psi_m\rangle = \int\limits_{-\infty}^\infty dX_m\, \psi_m (X_m; \theta)
|X_m (\theta)\rangle\;,
\label{eq:expan}
\end{equation}  
where $\psi_m (X_m;\theta)\equiv \langle X_m(\theta)|\psi_m\rangle$
denotes the wave function of the meter state at the angle $\theta$,
it is straightforward to find~\cite{kn:lou}
\begin{eqnarray}
 & \exp &[-i(2\sigma tX_s)\hat X_m(\phi)]|\psi_m \rangle
\nonumber \\
 & = & \int\limits_{-\infty}^\infty dX_m\exp[-i\gamma(X_s,X_m;\theta-\phi)]
\label{eq:action} \\
 & \times & \psi_m[X_m - 2\sigma tX_s\sin (\theta-\phi);\theta]
 |X_m(\theta)\rangle\;,
\nonumber
\end{eqnarray}
where
\begin{eqnarray}
\gamma(X_s, X_m; \theta-\phi) & \equiv & (\sigma t X_s)^2\sin[2(\theta-\phi)]
\nonumber \\
 & & + 2\sigma t X_s X_m \cos(\theta-\phi)\;.
\label{eq:gamma}
\end{eqnarray}

Hence the combined quantum state reads
\begin{eqnarray}
|\Psi(t)\rangle &=& \int\limits_{-\infty}^\infty dX_s
\int\limits_{-\infty}^\infty dX_m \psi_s(X_s; \phi + \pi/2)
\nonumber\\
& & \quad \quad \times \psi_m [X_m - 2\sigma tX_s \sin (\theta - \phi); \theta]
\nonumber\\
& & \quad \quad \times \exp[-i\gamma(X_s,X_m; \theta-\phi)]
\nonumber\\
& & \qquad \times |X_s (\phi + \pi/2)\rangle| X_m (\theta) \rangle\;.
\label{eq:combstate}
\end{eqnarray}
We note that due to the coupling between the meter and the signal via the 
Hamiltonian Eq.~(\ref{eq:cohami}), the meter wave function
$\psi_m (X_m;\theta)$ at the angle $\theta$ gets shifted by an amount
$\delta X_m \equiv 2\sigma tX_s \sin (\theta - \phi)$.

\section{Effect of the Meter Measurement on the Signal State}
\label{cond}

In the present section we shall show how a measurement of the meter influences
the state of the signal.
Let us first consider an arbitrary quadrature state of phase angle
$\theta$.

According to Eq.~(\ref{eq:combstate}) the conditioned state
\begin{equation}
|\psi_s^{(c)}\rangle=\frac{1}{\sqrt{W(X_m)}}\langle X_m(\theta)|\Psi(t)\rangle
\label{eq:condstate}
\end{equation}
of the signal given that our quadrature measurement at angle $\theta$ has
provided the value $X_m$ reads
\begin{equation}
|\psi_s^{(c)} \rangle = \int\limits_{-\infty}^\infty\, dX_s \psi_s (X_s; \phi
+ \pi/2) f (X_s|X_m)| X_s (\phi + \pi/2)\rangle\;,
\label{eq:estra}
\end{equation}
where the ``filter function'' $f$ is given by
\begin{eqnarray}
f (X_s| X_m) &=& \frac{1}{\sqrt{W(X_m)}} \psi_m [X_m - 2\sigma tX_s
\sin (\theta- \phi); \theta]
\nonumber\\
& \times & \exp[-i \gamma (X_s, X_m; \theta - \phi)]\;.
\label{eq:filt}
\end{eqnarray}
The normalization condition directly yields he probability 
$W(X_m)$ of finding the meter variable $X_m$, that is
\begin{eqnarray}
W (X_m) = \int\limits_{-\infty}^\infty & dX_s &|\psi_s (X_s; \phi + \pi/2)|^2
\nonumber\\
 & \times &|\psi_m[X_m - 2\sigma tX_s \sin (\theta - \phi); \theta]|^2\;.
\label{eq:dostar}
\end{eqnarray}
Equation (\ref{eq:estra}) clearly shows how the measurement of the meter
influences the quantum state of the signal: The filter function determined
by the wave function of the meter selects those parts of the signal wave
function that are entangled with the corresponding parts in the meter.
To study this in more detail we now calculate the Wigner function
\begin{eqnarray}
W_s^{(c)} (X_s, P_s| X_m) &=& \frac{1}{2\pi} \int\limits_{-\infty}^\infty dY 
\, e^{iP_sY}\langle X_s - Y/2| \psi_s^{(c)}\rangle
\nonumber\\ 
& & \qquad \quad \times \langle \psi_s^{(c)}|X_s + Y/2 \rangle
\label{eq:wigsigcon}
\end{eqnarray}
of the signal state conditioned on the measured meter value $X_m$.
Substituting the state $|\psi_s^{(c)}\rangle$, Eq.~(\ref{eq:estra}),
into this expression we arrive at
\begin{eqnarray}
&W&_s^{(c)}(X_s,P_s| X_m)=\frac{1}{2\pi} \int\limits_{-\infty}^\infty
dY \, e^{iP_sY} \psi_s (X_s -Y/2)
\nonumber \\
& \times &  \psi_s^\ast (X_s + Y/2) f(X_s -Y/2| X_m) f^\ast(X_s + Y/2| X_m)\;.
\label{eq:wigpar}
\end{eqnarray}
The integral may be expressed as the convolution
\begin{eqnarray}
& & W_s^{(c)} (X_s, P_s| X_m) =
\nonumber \\
& & \qquad \int\limits_{-\infty}^\infty dP'\, W_s (X_s, P_s - P')
W_f (X_s, P'| X_m)
\label{eq:conv}
\end{eqnarray}
between the Wigner function of the original signal state 
\begin{eqnarray}
& & W_s (X_s, P_s) = 
\nonumber \\
& & \qquad \frac{1}{2\pi} \int\limits_{-\infty}^\infty dY \, e^{iP_sY}
\psi_s (X_s -Y/2) \psi_s^\ast (X_s + Y/2)\;,
\label{eq:wig1}
\end{eqnarray}
and the Wigner function
\begin{eqnarray}
& & W_f (X_s, P_s| X_m) =
\nonumber \\
& &  \frac{1}{2\pi} \int\limits_{-\infty}^\infty dy \, e^{iP_sY}
f(X_s - Y/2| X_m) f^\ast (X_s + Y/2| X_m)
\label{eq:wig2}
\end{eqnarray}
of the ``filter'' provided by the measurement on the meter.

\section{Special Examples}
\label{exa}

\subsection{In phase measurement}
\label{inphase}

If we take the angle $\theta$ equal to 
$\phi$, the state $|\Psi \rangle$, Eq.~(\ref{eq:combstate}), of the complete
system reduces to
\begin{eqnarray}
|\Psi(t)\rangle &=& \int\limits_{-\infty}^\infty\, dX_s
\int\limits_{-\infty}^\infty dX_m\psi_s(X_s;\phi+\pi/2)\psi_m(X_m;\phi)
\nonumber \\
& & \times \exp(-i2\sigma tX_s X_m)|X_s(\phi + \pi/2)\rangle|X_m(\phi)
\rangle\;.
\label{eq:redstate}
\end{eqnarray}
In this case the meter wave function is not shifted.
Nevertheless, the states of signal and meter are still entangled.
Since the shift $\delta X_m$ vanishes, the probability  
\begin{equation}
W(X_m) = |\psi_m|^2 \int\limits_{-\infty}^\infty
         dX_s|\psi_s(X_s)|^2= |\psi_m|^2 \;,
\label{eq:prob}
\end{equation}
of finding the meter variable $X_m$ following from Eq.~(\ref{eq:dostar})
for $\theta = \phi$ is identical to the initial probability of the meter,
that is
\begin{equation}
W(X_m)=|\psi_m(X_m)|^2\;.
\label{eq:inprob}
\end{equation}
Hence, up to an overall phase $\mu_m$ determined by the meter
wave function $\psi(X_m)=|\psi(X_m)|\exp[i\mu(X_m)]$, we find from 
Eq.~(\ref{eq:filt}) the filter function $f(X_s|X_m)=\exp(-i2\sigma t X_s X_m)$,
and from Eq.~(\ref{eq:estra}) the conditioned signal state
\begin{equation}
|\psi_s^{(c)}\rangle = \int\limits_{-\infty}^{\infty}\, dX_s\psi_s(X_s)\exp
(-i2\sigma tX_s X_m)|X_s (\phi + \pi/2)\rangle\;.
\label{eq:stern}
\end{equation}
Note that the measurement of the meter has indeed changed the {\it state} of
the system but did not alter the probability 
\begin{eqnarray}
W(X_s) &=& |\langle X_s|\psi_s^{(c)}\rangle|^2
        = |\psi_s(X_s)\exp(-i2\sigma t X_s X_m)|^2
\nonumber\\
       &=& |\psi_s(X_s)|^2
\label{eq:probab}
\end{eqnarray}
of finding the signal variable $X_s$.
Hence, the measurement has left untouched the shape of the original state
(in the Wigner function representation) but has moved it along the momentum
axis by an amount of $2\sigma tX_m$. Consequently, the measurement did not
change the probability distribution in the conjugate variable, namely the
$X_s$ variable. We note, however, that in this way we cannot gain information
about the signal since according to Eqs.~(\ref{eq:prob}) and (\ref{eq:inprob})
the probability distribution $W(X_m)$ of measuring the variable $X_m$ is
identical to the original distribution.

This finding is actually a rather general result. In fact, it can be
rigorously shown~\cite{kn:noi} that a (QND) measurement which does not
change the probability density of the observable which is being measured
on a single quantum system gives no information about the measured observable.

\subsection{Out of phase measurement}
\label{outofphase}

Turning now to the case of $\theta = \phi + \pi/2$, we see that the shift
$\delta X_m = 2\sigma tX_s$ in the meter wave function is maximal and
according to Eq.~(\ref{eq:gamma}) the phase $\gamma$ vanishes. Hence, the
combined state
\begin{eqnarray}
&\null&|\Psi(t)\rangle = \int\limits_{-\infty}^\infty dX_s
\int\limits_{-\infty}^\infty dX_m \psi_s(X_s; \phi + \pi/2)
\label{eq:costate} \\
&\times&\!\psi_m (X_m-2\sigma tX_s;\phi + \pi/2)\,
          |X_s(\phi + \pi/2)\rangle\,|X_m(\phi + \pi/2)\rangle
\nonumber
\end{eqnarray}
is an entangled state in which the entanglement between the meter and signal
is due to the shift of the meter. In contrast to the discussion of
Sec.~\ref{inphase} we can now deduce properties of the signal from the shift
of the meter wave function. Unfortunately, we cannot simultaneously keep
the probability distribution $W(X_s)=|\psi_s(X_s)|^2$ of the original
signal state invariant, in accordance with the discussion at the end of
Sec.~\ref{inphase}.
Indeed, we find from Eqs.~(\ref{eq:estra}) or (\ref{eq:stern}) the conditional
state 
\begin{equation}
|\tilde{\psi}_s ^{(c)} \rangle = \frac{1}{\sqrt{\tilde{W}(X_m)}} \; \int
\limits_{-\infty}^\infty \!\!dX_s \, \psi_s (X_s) \psi_m (X_m - 2 \sigma t X_s)
\,|X_s \rangle
\label{eq:contilde}
\end{equation}
of the system given the meter measurement at phase $\phi+\pi/2$ has 
provided the value $X_m$. The probability
\begin{equation}
\tilde{W}(X_m) = \int\limits_{-\infty}^\infty dX_s |\psi_s (X_s) |^2|
\psi_m (X_m - 2 \sigma tX_s)|^2
\label{eq:wtilde}
\end{equation}
of finding the meter value $X_m$ following from Eq.~(\ref{eq:dostar}) is now a 
convolution of the system and the meter function.

\section{Special Measurements}
\label{meter}

\subsection{Weak measurements}
\label{weak}

If $\psi_m$ is broad compared to $\psi_s$ we can evaluate $\psi_m$ at 
some characteristic value of $X_s$, such as $\langle X_s \rangle$. As
a consequence the conditional state, Eq.~(\ref{eq:contilde}), is simply
given by
\begin{equation}
|\psi_s^{(c)} \rangle \equiv \int\limits_{-\infty}^{\infty}
dX_s\, \psi_s (X_s)| X_s \rangle\;.
\label{eq:weakcon}
\end{equation}
The probability 
\begin{equation}
\tilde{W}(X_m) \equiv |\psi_m (X_m+2\sigma t\langle X_s \rangle)|^2
\label{eq:weakprob}
\end{equation}
reduces to the original meter probability shifted by an amount
$2\sigma t\langle X_s\rangle$. Hence, when this shift
$2\sigma t\langle X_s \rangle$ is larger than the width of
$W_m (X_m) = |\psi_m (X_m)|^2$, we can 
learn about $\langle X_s \rangle$. As seen from Eq.~(\ref{eq:weakcon}),
in this case the state of the signal mode does not change appreciably.

\subsection{Tomographic measurements}
\label{tomo}

In the present section we show that it is possible to perform tomography
on the meter mode to obtain information about the signal state. To this
end, we rewrite Eq.~(\ref{eq:wtilde})
\begin{equation}
\tilde{W}(X_m) = \int\limits_{-\infty}^\infty dX_s |\psi_s (X_s) |^2|
\psi_m (X_m - 2 \sigma tX_s)|^2\;,
\label{eq:tildew}
\end{equation}
which gives the marginal distribution of the meter (probability distribution
of the results of the measurements of $\hat{X}_m$) in the case of out of
phase measurements. Let us assume that the meter wave function is extremely
narrow, that is the meter is initially in a highly squeezed state, for example
a squeezed vacuum $|0,r\rangle$, where $r$ is the (real) squeezing parameter.
Then, according to
Eq.~(\ref{eq:tildew}), the marginal distribution $\tilde{W}(X_m)$ is given by
a convolution of the modulus square of the signal wave function with a narrow
Gaussian
\begin{eqnarray}
|\psi_m(X_m&-&2\sigma tX_s)|^2=\frac{1}{\protect\sqrt{\pi}\cosh
r(1-\tanh r)}
\label{eq:psivac} \\
&\times&\exp\left\{-\left[\frac{1+\tanh r}{1-\tanh r}\right]
(X_m-2\sigma tX_s)^2\right\}\;.
\nonumber
\end{eqnarray}

Now, if the squeezing parameter $r$ is large enough, the
Gaussian~(\ref{eq:psivac}) approaches a delta function in the meter and
signal variables
\begin{equation}
|\psi_m(X_m-2\sigma tX_s)|^2\longrightarrow \frac{1}{|2\sigma t|}
\delta\left(X_s-\frac{X_m}{2\sigma t}\right)\;,
\label{eq:appdel}
\end{equation}
and Eq.~(\ref{eq:tildew}) reduces to
\begin{mathletters}
\label{eq:double}
\begin{eqnarray}
\tilde{W}(X_m)&\cong&\frac{1}{|2\sigma t|}\int\limits_{-\infty}^{\infty}
dX_s\, |\psi_s(X_s)|^2\delta\left(X_s-\frac{X_m}{2\sigma t}\right)
\label{eq:doublea} \\
&=& \frac{1}{2\sigma t} \left|\psi_s\left(\frac{X_m}{2\sigma t}\right)\right|^2
=\frac{1}{2\sigma t} W\left(\frac{X_m}{2\sigma t}\right)\;.
\label{eq:doubleb}
\end{eqnarray}
\end{mathletters}
Hence, by measuring the probability distribution $\tilde{W}(X_m)$ of the
outcomes of the meter variable $X_m$ (for example via balanced homodyne
detection performed on the meter field) we indirectly obtain the probability
distribution $W(X_s)$, up to a rescaling given by the factor $2\sigma t$.
However, from Eq.~(\ref{eq:contilde})
it is clear that in this case the signal wave function is changed, and
therefore we need to prepare the signal field again in the same state after
each measurement. This is what is usually done in quantum optical
tomography~\cite{kn:tomo}.

The advantage of the present scheme is that we perform an indirect measurement:
We do not detect the signal mode outside the cavity (that is, we do not have
to take the signal field outside the cavity), but we couple it to a
meter field which is successively detected, thus overcoming the smearing effect
introduced by the direct detection of the signal~\cite{kn:tomo}. Moreover, there
is no need of a smoothing procedure, since we are interested in the marginal
probability distribution $W(X_s)$ which is directly related to $\tilde{W}(X_m)$
through Eq.~(\ref{eq:double}). In order to probe the full state of the signal
field, however, we would need to measure the probability distribution
$\tilde{W}(X_m)$ for various values of the phase~\cite{kn:tomo}.

\section{Conclusions}
\label{conclu}

To summarize, we have suggested a method to measure the quadrature probability
distribution (or, more generally, the full quantum state) of a single mode
of the electromagnetic field inside a cavity. It is based on indirect
homodyne measurements performed on a meter field which is coupled to the
signal field via a QND interaction Hamiltonian. We have named this procedure
``endoscopic tomography'' because (i) it does not require (in contrast to
Ref.~\cite{kn:tomo}) to take the field out of the cavity, just as in
``quantum state endoscopy''~\cite{kn:bard}, where a beam of two-level atoms
is used as a probe; (ii) tomographic measurements performed (by balanced
homodyne detection) on the meter mode allow us to reconstruct the marginal
probability distribution of the signal variable or even the full quantum state.

\acknowledgments

It is a pleasure for us to acknowledge several enlightening discussions
with O.~Alter, F.~Harrison, J.H.~Kimble and K.~M\"olmer.
We thank the European Union (through the TMR Programme), the Deutsche
Forschungsgemeinschaft, the Land Baden-W\"urttemberg, and INFM
(through the 1997 PRA-CAT) for partial support.

\end{document}